\documentclass[aps,prd,floatfix,nofootinbib]{revtex4} % preprint, % ,nofootinbib, twocolumn, showpacs,groupedaddress,

\usepackage{graphics}
\usepackage{graphicx}

\usepackage{latexsym}
\usepackage[cp866]{inputenc}
\usepackage[english]{babel}

\input epsf

\begin{document}

\title{\boldmath Semileptonic decays $D\to\pi^+\pi^-e^+\nu_e$ and
$D_s\to\pi^+\pi^-e^+\nu_e$ as the probe of constituent
quark-antiquark pairs in the light scalar mesons}
\author{N.~N.~Achasov,$^{1}$\footnote{achasov@math.nsc.ru}
A.~V.~Kiselev,$^{1,2}$\footnote{kiselev@math.nsc.ru} and
G.~N.~Shestakov\,$^{1}$\footnote{shestako@math.nsc.ru}}
\affiliation{$^1$\,Laboratory of Theoretical Physics, S.~L.~Sobolev
Institute for Mathematics, 630090 Novosibirsk, Russia,
\\$^2$\,Novosibirsk State University, 630090 Novosibirsk, Russia}

%\date{}

\begin{abstract} Decays $D\to\pi^+\pi^-e^+\nu_e$ and $D_s\to\pi^+
\pi^-e^+\nu_e$ serve as probes that check the existence of
constituent $q\bar q$ components in the wave functions of scalar
mesons decaying into $\pi^+\pi^-$. There exists a great deal of
concrete evidence in favor of the exotic four-quark nature of light
scalars. At the same time, the further expansion of the area of the
$q^2\bar q^2$ model validity for light scalars on ever new processes
seems extremely interesting and important. We analyze the BESIII and
CLEO data on the decays $D^+\to \pi^+\pi^- e^+\nu_e$ and $D^+_s\to
\pi^+\pi^- e^+\nu_e$ and show that the results of these experiments
together can be interpreted in favor of the four-quark nature of
light scalar mesons $\sigma(500)$ and $f_0(980)$. Our approach can
also be applied to the description of other similar decays involving
light scalars.
\end{abstract}

\maketitle

\section{INTRODUCTION}

In the works \cite{AK12,AK14}, a program was proposed for studying
the $\sigma(500)$, $f_0(980)$, and $a_0(980)$ resonances in
semileptonic decays of $D$ and $B$ mesons. These decays provide
direct probe of constituent two-quark components in the wave
functions of light scalars \cite{AK12,AK14}. So for the decays of
$D^+_s$, $D^0$, and $D^+$ mesons we have: $D^+_s\to s\bar s\,e^+
\nu_e\to[\sigma(500)+f_0(980)]e^+\nu_e \to\pi^+\pi^-e^+\nu_e$,
$D^0\to d\bar u\,e^+\nu_e\to a^-_0(980) e^+\nu_e\to\pi^-\eta e^+
\nu_e$, $D^+\to d\bar d\,e^+ \nu_e\to a^0_0 (980)e^+\nu_e\to\pi^0
\eta e^+\nu_e$, and $D^+\to d\bar d\,e^+\nu_e \to[\sigma (500)+f_0
(980)]e^+\nu_e\to\pi^+\pi^-e^+\nu_e$. The development of this
program \cite{AK12, AK14,AK18,A20} resulted in evidences in favor of
the exotic nature of light scalar mesons. Certainly, there are many
theoretical works in which the semileptonic decays of $D$ mesons are
explored from many different aspects, see, for example, Refs.
\cite{WL10,Fa11,Ri12,Os15,So20} and references herein.

The available data on the branching fractions of the semileptonic
decays $D^+_s\to\pi^+\pi^- e^+\nu_e$ and $D^+\to\pi^+\pi^-e^+\nu_e$
involving light scalar mesons \cite{CLEO09,BESIII19, PDG2019} are
collected in Table 1. The CLEO and BESIII collaborations also
presented data on the shapes of the $\pi^+\pi^-$ $S$-wave mass
spectra in these decays \cite{CLEO09,BESIII19}.
\begin{table} [!ht]
\centering \caption{Branching fractions ($\mathcal{B}$) and widths
($\Gamma=\mathcal{B}/\tau_D$, where $\tau_D$ is the $D$ lifetime
\cite{PDG2019}) of semileptonic decays of the $D^+_s$ and $D^+$
mesons.}
\label{Tab1}\vspace*{0.1cm}%\begin{center}
\begin{tabular}{|l|c|c|c|}
 \hline
 Decay  & $\mathcal{B}$ $(\times10^{-4})$ & Collaboration & $\Gamma$ $(\times10^{8}s^{-1})$ \\ \hline
 $D^+_s\to f_0(980)e^+\nu_e,\  f_0(980)\to\pi^+\pi^-$   &  $20\pm3\pm1$ & CLEO  \cite{CLEO09}  & $39.7\pm6.3$\\
 $D^+ \to\sigma(500)e^+\nu_e,\  \sigma(500)\to\pi^+\pi^-$  &  $6.30\pm0.43\pm0.32$  & BESIII \cite{BESIII19} & $6.06\pm0.51$ \\
 $D^+  \to f_0(980)e^+\nu_e,\  f_0(980)\to\pi^+\pi^-$   &  $<0.28$ & BESIII \cite{BESIII19} & $<0.27$ \\
 \hline \end{tabular} %\end{center}
 \end{table} \vspace*{0.3cm} %\noindent
In this paper, in the light of the program \cite{AK12,AK14}, we
analyze the recent BESIII data \cite{BESIII19} on the decay
$D^+\to\pi^+ \pi^-e^+\nu_e$ together with the CLEO data
\cite{CLEO09} on the decay $D^+_s\to \pi^+\pi^- e^+\nu_e$. We show
that the results of these experiments on the $\pi^+\pi^-$ mass
spectra can be interpreted in favor of the four-quark nature of
light scalar mesons.

This paper is organized as follows. In Sec. II we present the
general formulas for the semileptonic decay widths of $D^+_s$ and
$D^+$ mesons into light scalars. In Sec. III we consider the
production of the mixed $\sigma(500)-f_0(980)$ resonance complex
which proceeds via direct couplings of $\sigma$ and $f_0$ with
$q\bar q$ pairs created in semileptonic decays of $D^+$ and $D^+_s$
mesons. We find a sharp contradiction of this production mechanism
with the data on the $\pi^+\pi^-$ mass spectra in the $D^+\to\pi^+
\pi^-e^+\nu_e$ decay. Section IV is devoted to an analysis of the
four-quark production mechanism of the $\sigma$ and $f_0$ states.
Within the existing data, this mechanism seems to be the most real.
This section also contains an important remark about the dip/peak
manifestation of the $f_0(980)$ resonance.

\section{Semileptonic decay widths}

\begin{figure} [!ht] % \vspace*{2mm}
\begin{center}\includegraphics[width=6.7cm]{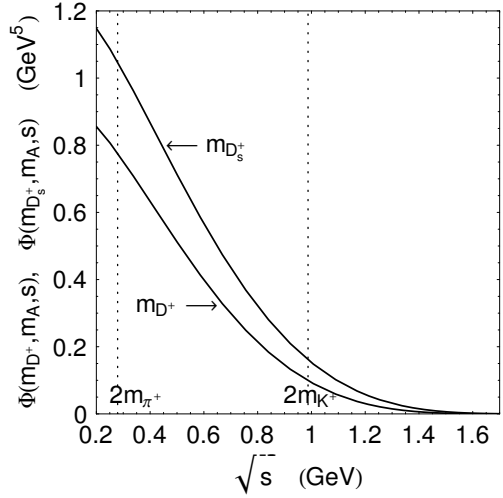}
\caption{\label{Fig1}The solid curves show the functions
$\Phi(m_{D^+},m_A,s)$ at $m_A=m_{D^+_{1}}=2.42$ GeV and
$\Phi(m_{D^+_s},m_A,s)$ at $m_A=m_{D^+_{s1}}=2.46$ GeV. The vertical
dotted lines indicate the $\pi^+\pi^-$ and the $K^+K^-$ threshold
positions.}\end{center}\end{figure}

First of all, we write the differential width for the $D^+$ and
$D^+_s$ decays into $\pi^+\pi^-e^+\nu_e$ in the form
\begin{eqnarray}\label{Eq1}\frac{d^2\Gamma_{D^+_{c\bar q}\to(S\to\pi^+\pi^-)e^+
\nu_e}(s,q^2)}{d\sqrt{s}\,dq^2} =\frac{G^2_F|V_{cq}|^2}{24\pi^3}
p^3_{\pi^+\pi^-}(m_{D^+_{c\bar q}},q^2,s)|f^{D^+_{c\bar
q}}_+(q^2)|^2 \frac{2\sqrt{s}}{\pi}|F^{D^+_{c\bar q}}_{q\bar q\to
S\to\pi^+ \pi^-}(s)|^2\rho_{\pi^+\pi^-}(s),\end{eqnarray} where the
index $q(\bar q)=d(\bar d),s(\bar s)$; $D^+_{c\bar d}\equiv D^+$,
$D^+_{c\bar s}\equiv D^+_s$, next we use the notation that is
convenient; $s$ and $q^2$ are the invariant mass squared of the
virtual scalar state $S$ (or the $\pi^+\pi^-$ system) and the
$e^+\nu_e$ system, respectively; $G_F$ is the Fermi constant,
$|V_{cq}|$ is a Cabibbo-Kobayshi-Maskawa matrix element (note that
$|V_{cs}|/|V_{cd}|\simeq20.92$ \cite{PDG2019}); $p_{\pi^+\pi^-}$ is
the magnitude of the three-momentum of the $\pi^+\pi^-$ system in
the $D$ meson rest frame,
\begin{eqnarray}\label{Eq2}
p_{\pi^+\pi^-}(m_{D^+_{c\bar q}},q^2,s)=\sqrt{[(m_{D^+_{c\bar
q}}-\sqrt{s})^2-q^2] [(m_{D^+_{c\bar q}}+
\sqrt{s})^2-q^2]}/(2m_{D^+_{c\bar q}} ),\end{eqnarray} and
$\rho_{\pi^+\pi^-}(s)=(1-4m^2_{\pi^+}/s)^{1/2}$. In a simplest pole
approximation, the form factor $f^{D^+_{c\bar q}}_+(q^2)$ has the
form \begin{eqnarray}\label{Eq3} f^{D^+_{c\bar
q}}_+(q^2)=\frac{f^{D^+_{c\bar q}}_+(0)}{1-q^2/m^2_A},\end{eqnarray}
where $m_A$, in principle, can be extracted from the data by fitting
\cite{CLEO09}. The amplitude $F^{D^+_{c\bar q}}_{q\bar q\to S\to
\pi^+\pi^-}(s)$ describes the formation and $\pi^+\pi^-$ decay of
the virtual scalar state $S$ produced in the $D^+_{c\bar
q}\to\pi^+\pi^-e^+\nu_e$ decay. For example, in case of direct
production of a single scalar resonance, $|F^{D^+_{c\bar q}}_{q\bar
q\to S\to \pi^+\pi^-}(s)|^2\rho_{\pi^+\pi^-}(s)=\sqrt{s}\Gamma_{S\to
\pi^+\pi^-}(s)/|D_S(s)|^2$, where $\Gamma_{S\to \pi^+\pi^-}(s)$ is
the $S\to\pi^+\pi^-$ decay width, $1/D_S(s)$ is the propagator of
$S$, and the amplitude normalization (in this case) is hidden in
$f^{D^+_{ c\bar q}}_+(0)$. The $\pi^+\pi^-$ invariant mass
distribution is given by
\begin{eqnarray}\label{Eq4}
\frac{d\Gamma_{D^+_{c\bar q}\to(S\to\pi^+\pi^-)e^+\nu_e}(s)}{
d\sqrt{s}} =\frac{G^2_F|V_{cq}|^2}{24\pi^3}|f^{D^+_{c\bar q}
}_+(0)|^2\,\Phi(m_{D^+_{c\bar q}},m_A,s)\frac{2\sqrt{s}}{\pi}
|F^{D^+_{c\bar q}}_{q\bar q\to S\to\pi^+\pi^-}(s)|^2\rho_{\pi^+
\pi^-}(s),\end{eqnarray} where
\begin{eqnarray}\label{Eq5}
\Phi(m_{D^+_{c\bar q}},m_A,s)=\int\limits_0^{(m_{D^+_{c\bar
q}}-\sqrt{s})^2}\frac{p^3_{\pi^+\pi^-}(m_{D^+_{c\bar q}},q^2,s)}
{|1-q^2/m^2_A|^2}\,dq^2.\end{eqnarray} Figure \ref{Fig1} illustrates
the energy dependence of $\Phi(m_{D^+_{c\bar q}},m_A,s)$ for $D^+$
and $D^+_s$ decays. Note that this function notably enhances the
$\pi^+\pi^-$ mass spectrum as $\sqrt{s}$ decreases.
%--------------------------------------------------------------------------------
%--------------------------------------------------------------------------------

\section{\boldmath $q\bar q$-probe in operation}

\begin{figure} [!ht] % \vspace*{2mm}
\begin{center}\includegraphics[width=12cm]{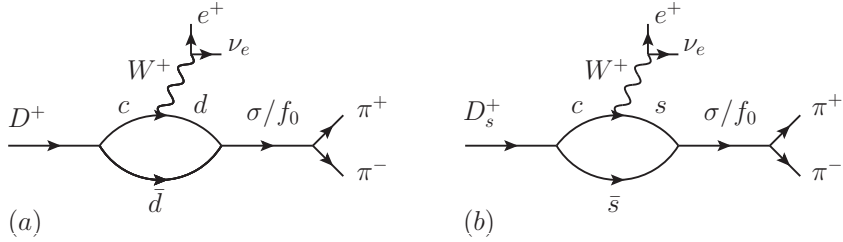}
\caption{\label{Fig2} Model of the $D^+\to(\sigma/f_0\to\pi^+\pi^-
)e^+\nu_e$ and $D^+_s\to(\sigma/f_0\to\pi^+\pi^-)e^+\nu_e$ decays.
}\end{center}\end{figure}

We now consider the production of the mixed $\sigma(500)-f_0(980)$
resonance complex (briefly $\sigma$ and $f_0$) which proceeds via
direct couplings of $\sigma$ and $f_0$ with $q\bar q$ pairs created
in semileptonic decays of $D^+$ and $D^+_s$ mesons (see Fig.
\ref{Fig2}).
%--------------------------------------------------------------------------------
%--------------------------------------------------------------------------------
This mechanism is the probe that verifies the existence of the
corresponding constituent $q\bar q$ component in the wave function
of a scalar meson. There exists a great deal of concrete evidence in
favor of the exotic four-quark nature of light scalars \cite{Ja77},
see also Ref. \cite{Wa13}. Reviews of the current situation can be
found, for example, in Refs. \cite{AK18,A20,AS11,AS19}. At the same
time, the further expansion of the area of the $q^2\bar q^2$ model
validity for light scalars on ever new processes seems to us
extremely interesting and important.

The transition amplitude $q\bar q\to S\to \pi^+\pi^-$ corresponding
to the indicated mechanism is denoted by $F^{D^+_{c\bar q},\,
direct}_{q\bar q\to S\to\pi^+\pi^-}(s)$ and write it in the form
\begin{eqnarray}\label{Eq6}F^{D^+_{c\bar q},\,direct}_{
q\bar q\to S\to\pi^+\pi^-}(s)=e^{i\delta^{\pi\pi}_B(s)}
\sum_{r,r'}g_{q\bar qr}\,G^{-1}_{rr'}\,g_{r'\pi^+\pi^-}=
e^{i\delta^{\pi\pi}_B(s)}\left(%
\begin{array}{cc}
g_{q\bar q\sigma}, & g_{q\bar qf_0}\\ \end{array}%
\right)\left(%
\begin{array}{cc} D_\sigma & -\Pi_{\sigma f_0}\\
-\Pi_{f_0\sigma} &  D_{f_0} \\ \end{array}%
\right)^{-1}\left(%
\begin{array}{c} g_{\sigma\pi^+\pi^-}\\
g_{f_0\pi^+\pi^-}\\ \end{array}%
\right),\end{eqnarray} where $r(r')=\sigma,f_0$; $\,g_{q\bar qr}$
and $g_{r\pi^+\pi^-}$ are the coupling constants, $D_r$ is the
inverse propagator of the unmixed scalar resonance $r$ with the mass
$m_r$\,, and $\Pi_{rr'}=\Pi_{r'r}$ is a nondiagonal element of the
polarization operator. $D_r$ has the form
\begin{eqnarray}\label{Eq7} D_r\equiv
D_r(s)=m^2_r-s+\sum_{ab}[\mbox{Re}\Pi^{ab}_r(m^2_r)-\Pi^{ab}_r(s)],
\end{eqnarray} where $\Pi^{ab}_r(s)$ stands for
the diagonal matrix element of the polarization operator of the
resonance $r$ corresponding to the contribution of the $ab$
intermediate state ($\pi^+\pi^-,\,\pi^0\pi^0$, $K^+K^-,\,K^0\bar
K^0$, etc). Re$\Pi^{ab}_r(s)$ is defined by the singly subtracted at
$s=0$ dispersion integral of
\begin{eqnarray}\label{Eq8}
\mbox{Im}\,\Pi^{ab}_r(s)=\sqrt{s}\Gamma_{r\to
ab}(s)=\eta_{ab}\frac{g^2_{r ab}}{16\pi}\rho_{ab}(s),\end{eqnarray}
where $g_{rab}$ is the coupling constant of $r$ with $ab$,\,
$\rho_{ab}(s)=\sqrt{s-m_{ab}^{(+)\,2}}\,\sqrt{ s-m_{ab}^{(-)
\,2}}/s$, $m_{ab}^{(\pm)}$\,=\,$m_a\pm m_b$ [here
$s>m_{ab}^{(+)\,2}$], and $\eta_{ab}=1$ ($1/2$) for different
(identical) decay particles $ab$, respectively. We also have
\begin{eqnarray}\label{Eq9}
\Pi_{rr'}\equiv\Pi_{rr'}(s)=C_{rr'}+\sum_{ab}\frac{g_{r' ab}}{g_{r
ab}}\Pi^{ab}_r(s),\end{eqnarray} where $C_{rr'}$ being the resonance
mixing parameter. The determinant of $G_{rr'}$ is $\Delta=D_\sigma
D_{f_0}-\Pi^2_{\sigma f_0}$. Thus the amplitudes for the $D^+$ and
$D^+_s$ decays have the form:
\begin{eqnarray}\label{Eq10} F^{D^+,\,direct}_{d\bar d\to
S\to\pi^+\pi^-}(s)=\frac{e^{i\delta^{\pi\pi}_B(s)}}{\Delta(s)}\left\{g_{d\bar
d\sigma}[D_{f_0}(s)g_{\sigma\pi^+\pi^-}+\Pi_{\sigma
f_0}(s)g_{f_0\pi^+\pi^-}]+ g_{d\bar df_0}[D_\sigma(s)
g_{f_0\pi^+\pi^-}+ \Pi_{\sigma f_0}(s)g_{\sigma\pi^+\pi^-}]\right\}.
\end{eqnarray}
\begin{eqnarray}\label{Eq11} F^{D^+_s,\,direct}_{s\bar s\to
S\to\pi^+\pi^-}(s)=\frac{e^{i\delta^{\pi\pi}_B(s)}}{\Delta(s)}\left\{g_{s\bar
s\sigma}[D_{f_0}(s)g_{\sigma\pi^+\pi^-}+\Pi_{\sigma
f_0}(s)g_{f_0\pi^+\pi^-}]+g_{s\bar sf_0}[D_\sigma(s)
g_{f_0\pi^+\pi^-}+ \Pi_{\sigma f_0}(s)g_{\sigma\pi^+\pi^-}]\right\}.
\end{eqnarray}
Here, we use the expressions and numbers from Ref. \cite{AK06}
(corresponding to fitting variant 1 from Table 1 therein) for
propagators $1/D_\sigma(s)$ and $1/D_{f_0}(s)$ of $\sigma(500)$ and
$f_0(980)$ resonances, the polarization operator matrix element
$\Pi_{\sigma f_0}(s)$, the $\delta^{\pi\pi}_B(s)$ phase of the
elastic background in the $S$-wave $\pi\pi$ scattering, $g_{\sigma
\pi^+\pi^-}$ and $g_{f_0\pi^+\pi^-}$ coupling constants, etc.

Note that our principal conclusions are independent of a concrete
fitting variants presented in Refs. \cite{AK06,AK11,AK12a},
containing the excellent simultaneous descriptions of the phase
shifts, inelasticity, and mass distributions in the reactions
$\pi\pi\to\pi\pi$, $\pi\pi\to K\bar K$, and $\phi\to\pi^0\pi^0
\gamma$. Also note that the expressions in square brackets in Eqs.
(\ref{Eq10}) and (\ref{Eq11}) are real for $\sqrt{s}$ below the
$K^+K^-$ threshold.

\begin{figure} [!ht] % \vspace*{2mm}
\begin{center}\includegraphics[width=13cm]{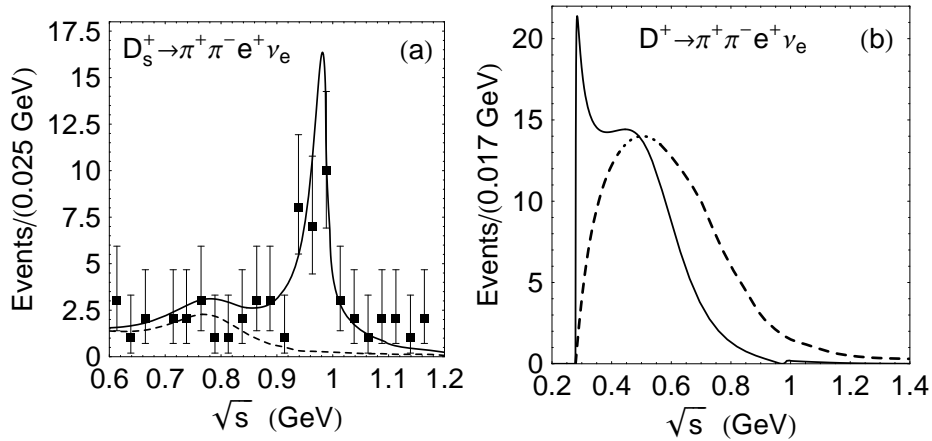}
\vspace*{-0.5cm}\caption{\label{Fig3}(a) The points with the error
bars are the CLEO data \cite{CLEO09} on the $\pi^+\pi^-$ invariant
mass distribution in the decay $D^+_s\to\pi^+\pi^-e^+\nu_e$
dominated by the $f_0(980)$ resonance production. The dashed curve
shows the total contribution from three noncoherent background
processes estimated by CLEO \cite{CLEO09}.
%--------------------------------------------------------------------------------
(b) The dashed curve represents the smoothed BESIII histogram with
0.017-GeV-wide-step for the $\pi^+\pi^-$ $S$-wave distribution
extracted by BESIII from the treatment of $D^+\to\pi^+ \pi^-e^+
\nu_e$ events \cite{BESIII19}. Uncertainties in the BESIII data can
range from $10\%$ to $20\%$. The $K^0_S$ veto region around 0.5 GeV
\cite{BESIII19} is shown by the dotted curve. The solid curves in
(a) and (b) correspond to the model described by Eqs.
(\ref{Eq10})--(\ref{Eq12}).}\end{center}\end{figure}

Consider the variant corresponding to the following simple choice of
direct coupling constants $\sigma$ and $f_0$ with $q\bar q$:
\begin{eqnarray}\label{Eq12}
g_{s\bar s\sigma}=0,\ \ \ g_{d\bar df_0}=0,\ \ \ g_{d\bar
d\sigma}=g_0/\sqrt{2},\ \ \ g_{s\bar sf_0}=g_0\,.
\end{eqnarray}
Further, without loss of generality, we put $g_0=1$. The
normalization constants $f^{D^+_s}_+(0)$ and $f^{D^+}_+(0)$ in
(\ref{Eq3}) are assumed to be equal. Then, substituting (\ref{Eq10})
and (\ref{Eq11}) into (\ref{Eq4}) and integrating over the intervals
$2m_\pi<\sqrt{s}<1.4$ GeV and $0.6$ GeV $<\sqrt{s}<1.2$ GeV,
respectively, we get the ratio of the widths
\begin{eqnarray}\label{Eq13} \frac{\Gamma_{D^+_s\to\pi^+\pi^-e^+\nu_e}
}{\Gamma_{D^+\to\pi^+\pi^-e^+\nu_e}}\approx5.62.\end{eqnarray} Thus,
we have satisfactory agreement with the data given in Table I,
according to which this ratio is equal to $6.55\pm1.18$. However,
Fig. \ref{Fig3} indicates that the joint description of the
$\pi^+\pi^-$ mass spectra in $D^+_s\to\pi^+\pi^-e^+\nu_e$ and
$D^+\to\pi^+\pi^- e^+\nu_e$ decays sharply contradicts the BESIII
\cite{BESIII19} data at $\sqrt{s}<1$ GeV. These data demonstrate a
smooth and wide $\pi^+\pi^-$ spectrum in the decay $D^+\to\pi^+\pi^-
e^+\nu_e$ [see Fig. \ref{Fig3}(b)], due to, according to the authors
of Ref. \cite{BESIII19}, the $\sigma(500)$ resonance production. It
is interesting that this contradiction is caused by the small mass
and large width of the unshielded $\sigma$ resonance
\cite{PDG2019,AK06,AK11,AK12a,AS94, AS07}, i.e., its main features.
The factor $\Phi(m_{D^+_{c\bar q}},m_A,s)$ in (\ref{Eq4}) more
enhances the $\pi^+\pi^-$ mass spectrum in the near-threshold region
(see Fig. \ref{Fig1}). Note that the fundamental role of the chiral
shielding in the fate of the $\sigma(500)$ meson was demonstrated in
the linear $\sigma$ model \cite{GL60} (which turned out to be a
nontrivial realization of QCD in the low-energy region) using
examples of the reactions $\pi\pi\to\pi\pi$ and $\gamma\gamma\to
\pi\pi$ \cite{AS94,AS07}.

But what is the sensitivity of the mass spectra shown in Fig. 3  to
possible deviations of $g_{s\bar s\sigma}$ and $g_{d\bar df_0}$ from
zero? Let the values of these constants are in the intervals:
\begin{eqnarray}\label{Eq12a}
-0.2<g_{s\bar s\sigma}<0.2,\ \ \ -0.2<g_{d\bar df_0}< 0.2 \nonumber
\end{eqnarray} [compare with Eq. (12) at $g_0=1$]. Then the ratio
$\Gamma_{D^+_s\to\pi^+\pi^-e^+\nu_e} /\Gamma_{D^+\to\pi^+\pi^-e^+
\nu_e}$ will be in the range from 5 to 7. From Eqs. (10) and (11) it
can be seen that the difference of $g_{s\bar s\sigma}$ from zero
affects only the amplitude $F^{D^+_s,\,direct}_{s\bar s\to S\to\pi^+
\pi^-}(s)$ and the difference of $g_{d \bar df_0}$ from zero affects
only the amplitude $F^{D^+,\,direct}_{d\bar d\to S\to\pi^+\pi^-}(s)
$. As a result, it turns out that the mass spectrum in Fig. 3(b)
varies slightly only in the $f_0(980)$ region. In most cases, the
expected small peak from $f_0(980)$ resonance appears in it. Thus, a
contradiction with the data presented in Fig. 3(b) remains
completely throughout the entire region $\sqrt{s}<1$ GeV. Difference
of $g_{s\bar s\sigma}$ from zero worsens the description of the
$\pi^+\pi^-$ mass spectrum in the decay $D^+_s\to\pi^+\pi^-e^+\nu_e$
in the $f_0(980)$ region shown in Fig. 3(a). Worsening is associated
with a noticeable rise of the left wing of the $f_0(980)$ resonance.
But a particularly significant effect of $\sigma(500)$ arises near
the $\pi^+\pi^-$ threshold when the $g_{s\bar s\sigma}\approx-0.2$.
The $\pi^+ \pi^-$ mass spectrum in the decay $D^+_s\to\pi^+\pi^-e^+
\nu_e$ at $\sqrt{s}<0.5$ GeV turns out to be similar to one in the
decay $D^+\to\pi^+\pi^-e^+\nu_e$ in the same region of $\sqrt{s}$
[see Fig. 3(b)]. Such a manifestation of the $\sigma(500)$ resonance
in $D^+_s\to\pi^+\pi^-e^+ \nu_e$ is extremely improbable.

So, we discard the above-described model of the creation of $\sigma$
and $f_0$ states due to the presence of $d\bar d$ and $s\bar s$
components in their wave functions, respectively. Figuratively, we
can say that the $q\bar q$ probe existing in semileptonic
$(D^+,D^+_s)\to\pi^+\pi^-e^+\nu_e$ decays does not find, to a first
approximation, the corresponding $q\bar q$ components.
%--------------------------------------------------------------------------------
%--------------------------------------------------------------------------------

It was directly shown in Ref. \cite{AK12} that the transition $s\bar
s\to\sigma(500)$ is negligible compared to the transition $s\bar
s\to f_0(980)$. In the work \cite{AK12}, it was also shown that the
intensity of the $s\bar s\to f_0(980)$ transition is about thirty
percent of the intensity of the $s\bar s\to\eta_s$ (where $\eta_s=
s\bar s$), $g^2_{s\bar sf_0}/g^2_{s\bar s\eta_s}\approx0.3$,
contrary expected equality of these intensities in the
chiral-symmetric models like the Nambu-Jona-Lasinio one. The above
analysis obviously supports the conclusion made in Ref. \cite{AK12}
that the decay $D^+_s\to \pi^+\pi^-e^+\nu_e$ testifies to the
previous conclusions about the dominant role of the four-quark
components in $\sigma(500)$ and $f_0(980)$ mesons.

%--------------------------------------------------------------------------------

\section{Four-quark production mechanism}

\begin{figure} [!ht] % \vspace*{2mm}
\begin{center}\includegraphics[width=12cm]{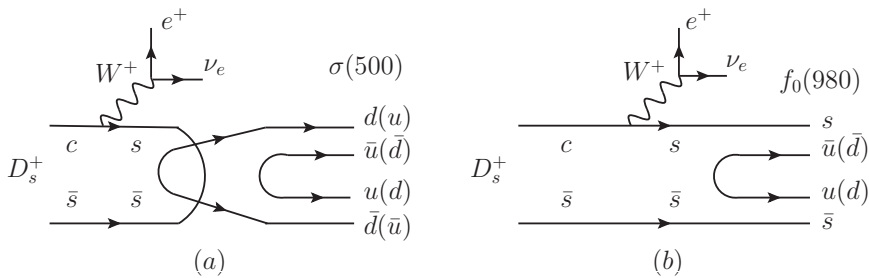}
\caption{\label{Fig4} Production of the four-quark $\sigma(500)$ and
$f_0(980)$ mesons in $D^+_s$ decays.}\end{center}\end{figure}

Let us now consider the four-quark $\sigma(500)=u\bar ud\bar d$ and
$f_0(980)=s\bar s(u\bar u+d\bar d)/\sqrt{2}$ meson production which
is symbolically depicted in the diagrams of Fig. 4 and 5. (We
emphasize that in the processing of the data we use, of course, the
resonance complex of the mixed states $\sigma$ and $f_0$ states
\cite{AK06,AK11,AK12a}.) These are ideal $q^2\bar q^2$ states of the
MIT bag with superallowed decays $\sigma\to\pi\pi$ and $f_0\to K\bar
K$ \cite{Ja77}. On the contrary, the decays $\sigma\to K\bar K$ and
$f_0\to\pi\pi$ are suppressed for these states by the
Okubo-Zweig-Iizuka (OZI) rule \cite{Ok63,Zw80, Ii66,Lip1,Lip2}. Due
to the small mass of $\sigma$, the OZI suppressed decay $\sigma\to
K\bar K$ does not play any role at all. At the same time, the main
decay of $f_0(980)$ under the $K\bar K$ threshold is precisely the
decay $f_0(980)\to\pi\pi$ due to a small $\sigma-f_0$ mixing. Thus,
the decay $D^+_s\to\pi^+\pi^-e^+\nu_e$, owing to the OZI-suppression
of the $\sigma$ resonance creation [see Fig. \ref{Fig4}(a)], is
dominated by the $f_0(980)$ resonance production [see Fig.
\ref{Fig4}(b)] followed by its decay into $\pi^+\pi^-$: $D^+_s\to
f_0(980)e^+\nu_e\to\pi^+ \pi^- e^+\nu_e$.
%--------------------------------------------------------------------------------
%--------------------------------------------------------------------------------
%--------------------------------------------------------------------------------
\begin{figure} [!ht] % \vspace*{2mm}
\begin{center}\includegraphics[width=12cm]{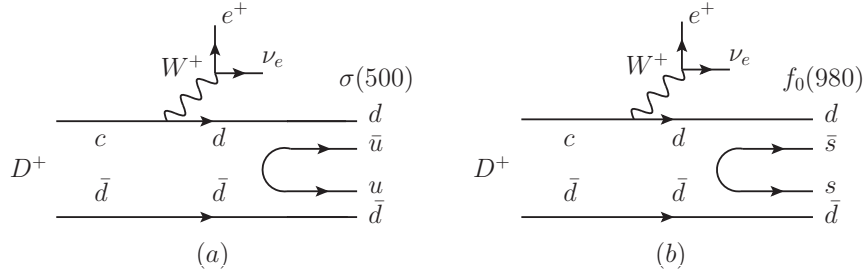}
\caption{\label{Fig5} Production of the four-quark $\sigma(500)$ and
$f_0(980)$ mesons in $D^+$ decays.}\end{center}\end{figure}
%--------------------------------------------------------------------------------

In the decay $D^+\to\pi^+\pi^-e^+\nu_e$, production of the
four-quark states $\sigma(500)$ and $f_0(980)$ is not suppressed by
the OZI rule, see Fig. \ref{Fig5}, and it would seem that both
states should manifest themselves as enhancements in the $\pi^+
\pi^-$ mass spectrum. However, the remarkable fact confirmed in many
reactions is that when there are no valence $s\bar s$ pairs in the
generating channel, the $f_0(980)$ resonance manifests itself (each
time) in the $\pi\pi$ mass spectrum not in the form of a peak, but
in the form of a sharp dip or sharp ledge, or a completely
insignificant fluctuation. The reason for this is the destructive
interference of the $f_0(980)$ contribution with a large and smooth
background, which is present in the $\pi\pi$ decay channel and has a
phase of $\approx90^\circ$. Striking examples here are the data on
the reactions $\pi\pi\to\pi\pi$ \cite{pipi1,pipi2}, $pp\to
p(\pi\pi)p$ \cite{ppipip}, $J/\psi\to\omega\pi^+\pi^-$ \cite{Omf0},
$\Upsilon(10860)\to\Upsilon(1S)\pi^+\pi^-$ \cite{Ypipi}, and, of
course, the discussed new BESIII data on $D^+\to\pi^+\pi^-e^+\nu_e$
\cite{BESIII19} (see also in this connection a comment in Ref.
\cite{FN2}).

And vice versa, when valence $s\bar s$ pairs are present in the
generating channel, such as in the reactions
$K^-p\to\pi^+\pi^-(\Lambda, \Sigma^0)$ \cite{Kp}, $J/\psi\to\phi
\pi^+\pi^-$ \cite{Phif0}, $D^+_s\to\pi^+\pi^+\pi^-$ \cite{Ds3pi},
and $D^+_s\to\pi^+\pi^-e^+\nu_e$ \cite{CLEO09}, then a sharp peak is
observed in the $f_0(980)$ resonance region.

\begin{figure} [!ht] % \vspace*{2mm}
\begin{center}\includegraphics[width=16cm]{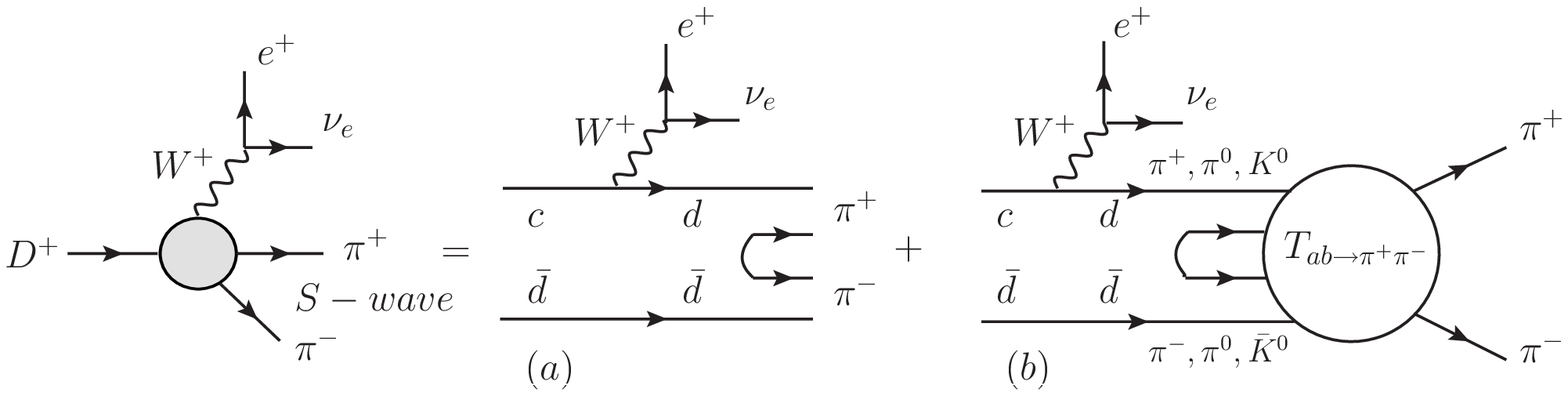}
\caption{\label{Fig6} The semileptonic decay
$D^+\to\pi^+\pi^-e^+\nu_e$ decays.}\end{center}\end{figure}

\begin{figure} [!ht] % \vspace*{2mm}
\begin{center}\includegraphics[width=12cm]{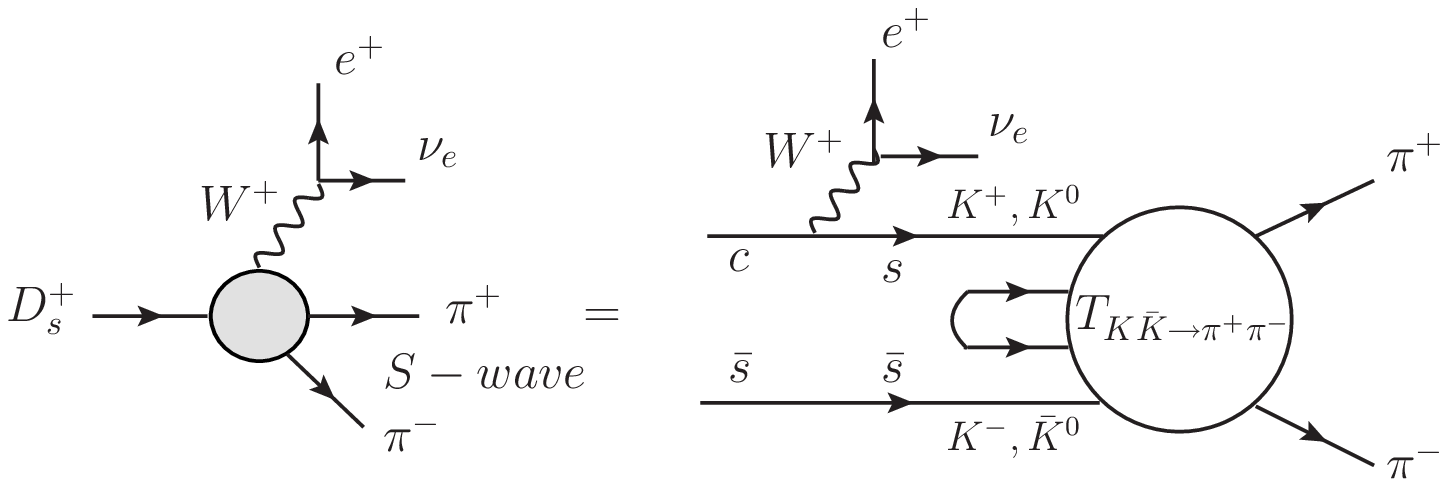}
\caption{\label{Fig7} The semileptonic decay
$D^+_s\to\pi^+\pi^-e^+\nu_e$ decays.}\end{center}\end{figure}

The described picture of the creation of four-quark resonances in
the $D^+\to\pi^+\pi^-e^+\nu_e$ and $D^+_s\to\pi^+\pi^-e^+\nu_e$
decays can be effectively realized in the language of hadronic
states, see Figs. \ref{Fig6} and \ref{Fig7}.
%--------------------------------------------------------------------------------
%--------------------------------------------------------------------------------
%--------------------------------------------------------------------------------
%--------------------------------------------------------------------------------
The mechanisms indicated in Figs. \ref{Fig6} and \ref{Fig7} imply
that the $S$-wave $\pi^+\pi^-$ system can be produced via seed
four-quark fluctuations $d\bar d\to\pi\pi$, $d\bar d\to K\bar K$,
and $s\bar s\to K\bar K$, which are then dressed by strong
interactions in the final state. According to Figs. \ref{Fig6} and
\ref{Fig7}, we write the amplitudes $F^{D^+}_{d \bar d\to
S\to\pi^+\pi^-}(s)$ and $F^{D^+_s}_{s \bar s\to S\to\pi^+\pi^-}(s)$
from Eq. (\ref{Fig4}) in the form
\begin{eqnarray}\label{Eq14}
F^{D^+ }_{d \bar d\to S\to\pi^+\pi^-}(s)=\lambda_{d\bar d\pi^+
\pi^-}\left[1+I_{\pi^+\pi^-}(s) \,T^0_0(s)\right]+\lambda_{d\bar
dK^0\bar K^0}I_{K^0\bar K^0}(s)\,T_{K^0 \bar K^0\to\pi^+\pi^-}(s),
\end{eqnarray}
\begin{eqnarray}\label{Eq15}
F^{D^+s}_{s \bar s\to S\to\pi^+\pi^-}(s)=\lambda_{s\bar sK^0\bar
K^0}\left[I_{K^+K^-}(s)+I_{K^0\bar K^0}(s)\right]\,T_{K^0 \bar
K^0\to\pi^+\pi^-}(s), \end{eqnarray} where
$T^0_0(s)=T_{\pi^+\pi^-\to\pi^+\pi^-}(s)+\frac{1}{2}T_{\pi^0\pi^0
\to\pi^+\pi^-}(s)$ is the $S$-wave amplitude of the reaction
$\pi\pi\to\pi\pi$ in the channel with isospin $I=0$ composed of the
amplitudes related to individual charge channels;
$T^0_0(s)=[\eta^0_0(s)\exp(2i\delta^0_0(s))-1]/(2i\rho_{\pi^+\pi^-}(s))$,
where $\eta^0_0(s)$ and $\delta^0_0(s)$ are the corresponding
inelasticity and phase of $\pi\pi$ scattering; $T_{K^0\bar
K^0\to\pi^+ \pi^-}(s)$ is the amplitude of the $S$-wave transition
$K^0\bar K^0 \to\pi^+\pi^-$; $T_{K^+ K^-\to\pi^+\pi^-}(s)=T_{K^0
\bar K^0 \to\pi^+\pi^-}(s)$ \cite{AK06,AK11,AK12a,FN1,CCL06}.
Functions $I_{a\bar a}(s)$ (where $a\bar a=\pi^+\pi^-,K^+K^-,K^0\bar
K^0$) are the amplitudes of the one-loop two-point diagrams
describing $a\bar a\to a\bar a\to$({\it the scalar state with a mass
equaling} $\sqrt{s}$) transitions in which initial $a\bar a$ pairs
are produced by $q\bar q$ sources described by coupling constants
$\lambda_{q\bar qa\bar a}$. Above the $a\bar a$ threshold, $I_{a\bar
a}(s)$ has the form \cite{AK06}
\begin{eqnarray}\label{Eq16}
I_{a\bar a}(s)=\tilde{C}_{a\bar a}+\rho_{a\bar
a}(s)\left(i+\frac{1}{\pi}\ln\frac{1+\rho_{a\bar a}(s)
}{1-\rho_{a\bar a}(s)}\right),\end{eqnarray} where $\rho_{a\bar
a}(s)=\sqrt{1-4m^2_a/s}$ (we put $m_{\pi^0}=m_{\pi^+}$ and take into
account the mass difference of $K^+$ and $K^0$); if $\sqrt{s}<2m_K$,
then $\rho_{K\bar K}(s)\to i|\rho_{K\bar K}(s)|$;
$\tilde{C}_{\pi^+\pi^-}$ and $\tilde{C}_{K^+K^-}=\tilde{C}_{K^0 \bar
K^0}$ are subtraction constants in the loops.

For reasons of $SU(3)$ symmetry, we will assume that all seed
coupling constants in Eqs. (\ref{Eq14}) and (\ref{Eq15}) are the
same: $\lambda_{d\bar d\pi^+\pi^-} =\lambda_{s \bar sK^0\bar
K^0}=\lambda_{s\bar sK^+K^-}= \lambda_{d\bar dK^0\bar K^0}$. For
reasons of $SU(4)$ symmetry, $f^{D^+_s}_+(0)=f^{D^+}_+ (0)$. Then,
for example, the product $f^{D^+_s}_+(0)\lambda_{s\bar sK^0\bar
K^0}$ will determine the absolute normalization of the widths
$\Gamma_{D^+_s \to\pi^+\pi^- e^+\nu_e}$ and $ \Gamma_{D^+\to\pi^+
\pi^-e^+\nu_e}$. But the ratio $\Gamma_{D^+_s\to\pi^+\pi^-
e^+\nu_e}/\Gamma_{D^+\to \pi^+\pi^- e^+\nu_e}$ does not depend on
this parameter.

Since the amplitudes $T^0_0(s)$ and $T_{K^0\bar K^0\to\pi^+
\pi^-}(s)$ are known \cite{AK06,AK11,AK12a} from the analysis of the
data on the reactions $\pi\pi\to\pi\pi$, $\pi\pi\to K\bar K$, and
$\phi\to\pi^0\pi^0\gamma$, then we have only two parameters
$\tilde{C}_{\pi^+\pi^-}$ and $\tilde{C}_{K^+K^-}$ to describe the
$\pi^+\pi^-$ mass spectra in the decays $D^+\to\pi^+\pi^-e^+ \nu_e$
and $D^+_s\to\pi^+\pi^-e^+\nu_e$ as well as the value of the ration
$\Gamma_{ D^+_s\to\pi^+\pi^-e^+\nu_e}/\Gamma_{D^+\to \pi^+\pi^-e^+
\nu_e}$ in agreement with experiment.

\begin{figure} [!ht] % \vspace*{2mm}
\begin{center}\includegraphics[width=13cm]{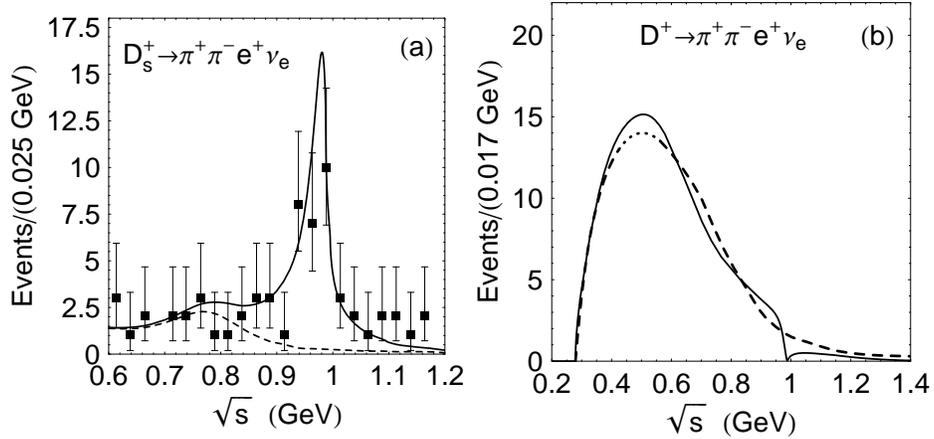}
\vspace*{-0.5cm}\caption{\label{Fig8} The same as in plots (a) and
(b) in Fig. \ref{Fig3}, but the solid theoretical curves correspond
to the model describable by Eqs. (\ref{Eq14})--(\ref{Eq16}).
}\end{center}\end{figure}

The choice of $\tilde{C}_{\pi^+\pi^-}=1.8$ and $\tilde{C}_{K^+K^-}
=1.0$ provides a good simultaneous description of the $\pi^+\pi^-$
mass spectra in the decays $D^+\to\pi^+ \pi^-e^+\nu_e$ and
$D^+_s\to\pi^+\pi^- e^+\nu_e$, see Fig. \ref{Fig8}, and gives the
ratio $\Gamma_{ D^+_s\to\pi^+ \pi^-e^+\nu_e}/\Gamma_{D^+\to\pi^+
\pi^-e^+\nu_e}\simeq6.55$, which is in excellent agreement with the
data. Let us note that Fig. \ref{Fig3}(b) demonstrates a sharp
contradiction with the BESIII data in all region of $\sqrt{s}$ for
the $q\bar q$ production mechanism, which is discussed immediately
below Eq. (13). In contrast, Fig. \ref{Fig8}(b) shows a good
agreement with the data in the case of the creation of four-quark
resonances.

%--------------------------------------------------------------------------------
%--------------------------------------------------------------------------------

In summary, in the light of the program \cite{AK12,AK14}, we have
analyzed the recent BESIII data \cite{BESIII19} on the decay
$D^+\to\pi^+ \pi^-e^+\nu_e$ together with the CLEO data
\cite{CLEO09} on the decay $D^+_s\to\pi^+\pi^-e^+\nu_e$ and showed
that the results on the $\pi^+\pi^-$ mass spectra of these
experiments together can be interpreted in favor of the four-quark
nature of light scalar mesons $\sigma(500)$ and $f_0(980)$. Our
approach can also be applied to the description of other similar
decays involving light scalars.\\

\begin{center}{\bf Acknowledgments}\end{center}

The study was carried out within the framework of the state contract
of the Sobolev Institute of Mathematics, Project No. 0314-2019-0021.

%--------------------------------------------------------------------------------


\begin{thebibliography}{99}
\bibitem{FN2} In Ref. \cite{WL10}, it is argued that
\begin{eqnarray}\label{Fn} R=\frac{\mathcal{B}(D^+\to
f_0(980)l^+ \nu)+\mathcal{B}(D^+\to
f_0(600)l^+\nu)}{\mathcal{B}(D^+\to
a^0_0(980)l^+\nu)}=\left\{\begin{array}{ll}1\quad\mbox{for two-quark scalars},\\
3\quad\mbox{for four-quark scalars}.
\end{array}\right.\nonumber\end{eqnarray} However, the BESIII
experiment \cite{BESIII19} suggests that $\mathcal{B}(D^+\to
f_0(980)e^+\nu_e\to\pi^+\pi^-e^+\nu_e)\approx0$. Thus, it becomes
clear that the naive quark counting rules cannot be applied in this
case for the reliable selection of the models for scalar mesons.
\bibitem{FN1} The amplitudes $T^0_0(s)$ and $T_{\pi\pi\to K\bar K}(s)$
were described in Refs. \cite{AK06,AK11,AK12a} by the complex of the
mixed $\sigma(500)$ and $f_0(980)$ resonances and smooth background
contributions using formulas similar to Eqs. (\ref{Eq10}) and
(\ref{Eq11}). The authors constructed the $\pi\pi$ scattering
amplitude $T^0_0$(s) \cite{AK06, AK11,AK12a} with regular analytical
properties in the $s$ complex plane, describing both experimental
data and the results based on chiral expansion and Roy equations
\cite{CCL06}. Note that the phases of the amplitudes $F^{D^+}_{d
\bar d\to S\to\pi^+\pi^-}(s)$ and $F^{D^+s}_{s\bar s\to S\to\pi^+
\pi^-}(s)$ in Eqs. (\ref{Eq14}) and (\ref{Eq15}) [taking into
account Eq. (\ref{Eq16})] coincide with the $\pi\pi$ scattering
phase $\delta^0_0(s)$ below the $K^+ K^-$ threshold where
$\eta^0_0(s)=1$.
\end{thebibliography}
\end{document}